\documentclass[aps,pra,superscriptaddress,twocolumn,showpacs]{revtex4}
\usepackage[colorlinks, linkcolor=blue, citecolor=blue, urlcolor=blue, breaklinks=true]{hyperref}
\usepackage{amsmath}
\usepackage{graphicx}
\usepackage{amssymb}
\usepackage{bbm, color}
\usepackage{relsize}
\usepackage[title]{appendix}
\usepackage{amsthm}
\usepackage{lipsum}
\usepackage{listings}
\usepackage{bbold}

\theoremstyle{plain}

\theoremstyle{plain}

\newcommand*{\cl}[1]{{\mathcal{#1}}}

\newcommand{\proj}[2]{| #1 \rangle\!\langle #2 |}

\newcommand{\ba}{\begin{eqnarray}}
\newcommand{\ea}{\end{eqnarray}}

\newcommand{\eq}[1]{(\hyperref[eq:#1]{\ref*{eq:#1}})}

\newcommand{\lemm}[1]{\hyperref[lemm:#1]{Lemma~\ref*{lemm:#1}}}

\DeclareMathOperator{\Tr}{Tr}

\begin{document}

\title{Upper bounds on the quantum capacity for general attenuator and amplifier}
\author{Youngrong Lim} 
\affiliation{School of Computational Sciences, Korea institute for Advanced Study, Seoul 02455, Korea}
\author{Soojoon Lee} 
\affiliation{Department of Mathematics and Research Institute for Basic Sciences, Kyung Hee University, Seoul 02447, Korea}
\affiliation{School of Computational Sciences, Korea institute for Advanced Study, Seoul 02455, Korea}
\author{Jaewan Kim} 
\affiliation{School of Computational Sciences, Korea institute for Advanced Study, Seoul 02455, Korea}
\author{Kabgyun Jeong}
\email{kgjeong6@snu.ac.kr}
\affiliation{Research Institute of Mathematics, Seoul National University, Seoul 08826, Korea}
\affiliation{School of Computational Sciences, Korea institute for Advanced Study, Seoul 02455, Korea}

\date{\today}

\begin{abstract}
There have been several upper bounds on the quantum capacity of the single-mode Gaussian channels with thermal noise, such as thermal attenuator and amplifier. We consider a class of attenuator and amplifier with more general noises, including squeezing or even non-Gaussian one. We derive new upper bounds on the energy-constrained quantum capacity of those channels by using the quantum conditional entropy power inequality. Also, we obtain lower bounds for the same channels by means of Gaussian optimizer with fixed input entropy. They give narrow bounds when the transmissivity is near unity and the energy of input state is low.
 \end{abstract} 

\maketitle
\section{introduction}

Quantum technology uses quantum phenomena like nonlocality and entanglement, in order to overcome the classical limitations in many areas such as quantum metrology, quantum computation, and simulation~\cite{roadmap}. Quantum communication is a significant area using quantum technology in which we expect critical advantages over classical communication with classical resources~\cite{Wilde}.  

Quantum capacity is a quantity measuring the ability to transmit quantum information, i.e., qubits, via a given quantum channel. In other words, it is the maximum achievable rate in the limit of infinitely many channel uses and vanishing error for the presence of noises in the channel.  We need to investigate the regularization of coherent information, which quantifies the quantum capacity of the channel~\cite{Lloyd, Devetak}. However, this quantity is hard to compute in general, owing to its non-additivity~\cite{DiVin, Smith, Cubitt}.

Bosonic Gaussian channels have been well studied because they can be implemented by simple quantum optical elements~\cite{RMP, Serafini}, such as beam splitter, phase shifter, and squeezer. Although the bosonic Gaussian channels are particular kinds of the generic quantum channels on continuous-variables, there still have interesting non-additive features for the quantum capacity called superactivation~\cite{Smith11} and activation~\cite{Lim18, Lim19}. The pure loss channel, a special kind of general Gaussian attenuators, can be described by a beam splitter mixing vacuum state with the input state, whose quantum capacity has been known precisely, as in the case of quantum-limited amplifier~\cite{Holevo,Wolf07}. There exist the thermal attenuator and the amplifier as more general cases. As their environment, the two channels use a thermal state instead of the vacuum state. For each of the channels, the exact value of quantum capacity of the channel has not been known, but only lower and upper bounds on the quantum capacity have been known by means of several methods~\cite{nohlower,upper,noh,plob,constrained}.

Quantum entropy power inequality (QEPI) is one of the useful tools in quantum information theory, firstly introduced by K\"{o}nig and Smith~\cite{KSIEEE,EPI}.
It tells us that the output entropy power does not decrease via the quantum mixing operation, e.g., a beam splitter, with two independent input states. QEPI has been proved recently and extended to the conditional cases~\cite{QEPI, cQEPI,cQEPI2,cQEPI3}. It is directly related with the bound on the minimum output entropy of given channels, then we can get the upper bounds on the classical information capacity. One of the advantages of using QEPI is that it is only dealing with the entropy values of quantum states, not details of the state itself. Consequently, it has been known that QEPI is applicable to general Gaussian noises and even non-Gaussian channels for obtaining upper bounds on the classical capacity of the channels~\cite{Konig,Jeong}. 

In this work, we apply the conditional quantum entropy power inequality (\textsc{cQEPI}) to general attenuator and amplifier channels, in which the environment can be general Gaussian states or even non-Gaussian states, in order to obtain new upper bounds on the quantum capacity for those channels. It is not only the first attempt to get a meaningful result on quantum capacity using QEPI, but also gives us an intuition that if more photons are in the environment of such a channel, then the channel has higher upper bound on the quantum capacity.

This paper is organized as follows: In Section~\ref{pre}, we introduce backgrounds to understand our results, and present new upper bounds on the quantum capacity for general attenuator and amplifier in Section~\ref{upper bounds}. In Section~\ref{lower bounds}, we derive the lower bounds as well and compare them with our upper bounds. Also, we give specific examples in Section~\ref{examples}, in order to present physical relevance. Finally, in Section~\ref{discussion}, we summarize our results, and comment on a few remarks and open problems.

\section{preliminaries}\label{pre}

\begin{figure}[!t]
\includegraphics[width=8.5cm]{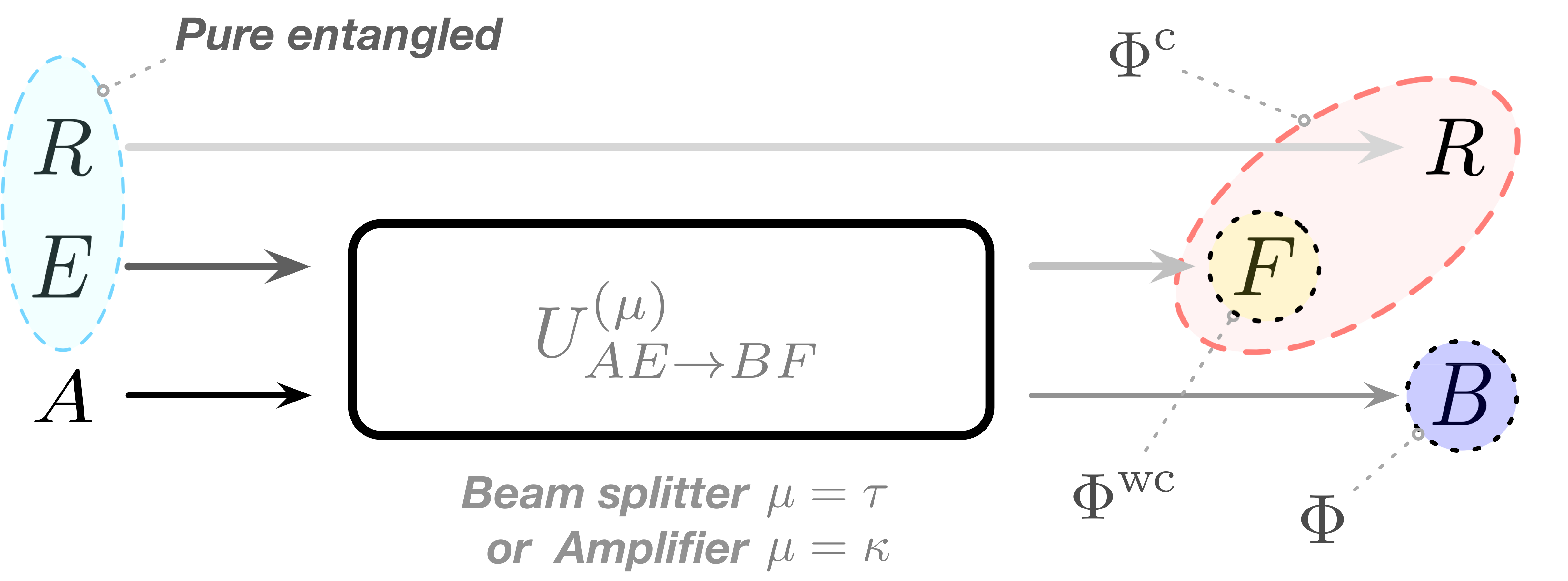}
\caption{A schematic diagram for complementary and weak-complementary Gaussian channels. For a mixed state of environment for a given channel, the purifying system is represented as $R$. } 
\label{fig1}
\end{figure}

The Stinespring dilation for a Gaussian quantum channel $\Phi$ can be written as
\begin{equation}
\Phi(\rho_A)=\text{Tr}_E\left[ U_{\Sigma}(\rho_A \otimes \rho_E)U^{\dagger}_{\Sigma}\right],
\end{equation}
where $U_{\Sigma}$ is a symplectic unitary transformation on the total Hilbert space $\cl{H}(L^2(\mathbb{R}^{n_A}))\otimes \cl{H}(L^2(\mathbb{R}^{n_E}))$ with the number of input mode $n_A$ and environment mode $n_E$, the Hilbert space of square integrable function $\cl{H}(L^2(\mathbb{R}^n))$ and $\rho_E$ denotes a pure Gaussian state (Notice that, in Fig.~\ref{fig1}, $U_{\Sigma}\equiv U_{AE\to BF}^{(\mu)}$). In this work, we deal with two important unitary operations given by
\begin{align}
&U_\lambda=\exp \left[ \arctan \sqrt{1-\lambda \over \lambda}(a^\dagger e-e^\dagger a)\right],\nonumber \\
&U_\kappa=\exp \left[ \text{arctanh} \sqrt{\kappa-1 \over \kappa}(a^\dagger e^{\mathsmaller{*}} -e^T a)\right],
\end{align}
where $\lambda \in [0,1]$, $\kappa \in [1,\infty]$, and $a$, $e$ are annihilation operators of input and environment, respectively. These are nothing but beam splitter with transmissivity $\lambda$ and amplifier with gain $\kappa$.
 
By Stinespring dilation, we can naturally define the complementary channel as
\begin{equation}\label{compure}
\Phi^\text{c}(\rho_A)=\text{Tr}_B\left[ U_{\Sigma}(\rho_A \otimes \rho_E)U^{\dagger}_{\Sigma}\right].
\end{equation}
However, if the environment state $\rho_E$ is mixed state, we cannot obtain the complementary channel uniquely by this method. Instead, firstly we need to purify the environment state, and then find the corresponding symplectic unitary in the extended Hilbert space. It can be expressed as 
\begin{equation}\label{com}
\Phi^\text{c}(\rho_A)=\text{Tr}_{B}\left[U_{\Sigma} \otimes \mathbb{1}_R (\rho_A\otimes \proj{\psi}{\psi}_{ER})(U_{\Sigma} \otimes \mathbb{1}_R)^{\dagger}\right],
\end{equation}
where $\proj{\psi}{\psi}_{ER}$ is a quantum purification such that $\text{Tr}_{R}\proj{\psi}{\psi}_{ER}=\rho_E$. Also, we can define weak-complementary channel $\Phi^\text{wc}$ as the case for which a mixed state $\rho_E$ is inserted in Eq.~(\ref{compure}), and $\Phi^\text{wc}=\Phi^\text{c}$ when $\rho_E$ is pure~\cite{Caruso06}. In Fig.~\ref{fig1}, we describe the situation, in which the environment is a non-pure state.

The quantum capacity of a channel $\Phi$ under a constraint with input mean photon number $N$ is defined by
\begin{equation}
{\cal Q}(\Phi,N)=\lim_{n \rightarrow \infty}\max_{\bar{E}(\rho_n) \le nN} \frac{I_c\left( \Phi^{\otimes n},\rho_n \right)}{n},
\end{equation}
where $\Phi^{\otimes n}$ is the $n$ independent uses of the channel, $\bar{E}(\rho_n)$ is energy of the input state, and $\rho_n$ is any input state in $n$ tensor product of the original Hilbert space of the input state for the single channel. The coherent information of a channel $\Phi$ and an input state $\rho_A \in \cl{H}(L^2(\mathbb{R}^{n_A}))$ can be written as
\begin{equation}
I_c(\Phi,\rho_A)=S\left(\Phi(\rho_A)\right)-S\left(\Phi^\text{c}(\rho_A)\right),
\end{equation}
where $S(\varrho)=-\Tr \varrho \log \varrho$ is the von Neumann entropy.

The linear version~\cite{KSIEEE} of QEPI is described as
\begin{equation}
S(\rho_{X_1}\boxplus_\tau\rho_{X_2})\ge\tau S(\rho_{X_1})+(1-\tau)S(\rho_{X_2}),
\end{equation}
where $\rho_{X_1}$ and $\rho_{X_2}$ are independent input states, and $\boxplus_\tau$ means a beam splitter operation with transmissivity $\tau \in [0,0.5]$.

\section{upper bounds on the quantum capacity }\label{upper bounds}
Now, we can think a general attenuator $\Phi_{\tau,\rho_E}$, in which the environment can be any Gaussian state or even non-Gaussian state. Then we can get an upper bound of this channel $\Phi_{\tau,\rho_E}$ as follows:
\begin{align}\label{Qcap}
&{\cal Q}(\Phi_{\tau,\rho_E},N):=\lim_{n \rightarrow \infty}\max_{\bar{E}(\rho_n) \le nN}{1 \over n} I_c\left( \Phi^{ \otimes n}_{\tau,\rho_E},\rho_n \right)  \nonumber\\
&=\lim_{n \rightarrow \infty}\max_{\bar{E}(\rho_n) \le nN} {1 \over n} \left[S( \Phi^{ \otimes n}_{\tau,\rho_E}(\rho_n))-S( \Phi^{c \otimes n}_{\tau,\rho_E}(\rho_n)) \right] \nonumber \\
&\le \lim_{n \rightarrow \infty} \max_{\rho_n}  {1 \over n}S( \Phi^{ \otimes n}_{\tau,\rho_E}(\rho_n))-\lim_{n \rightarrow \infty} \min_{\rho_n}  {1 \over n}S( \Phi^{c \otimes n}_{\tau,\rho_E}(\rho_n)) \nonumber \\
&\le \max_{\rho}S \left(\Phi_{\tau,\rho_E}(\rho) \right)-\lim_{n \rightarrow \infty} \min_{\rho_n}  {1 \over n}S( \Phi^{c \otimes n}_{\tau,\rho_E}(\rho_n)),
\end{align}
where the last inequality comes from the subadditivity of entropy. We know the upper bound of the first term of Eq.~(\ref{Qcap}) from the fact that Gaussian states always have maximal entropies for given first and second moments~\cite{Wolf}. Explicitly, we have 
\begin{equation}
\max_{\rho}S(\Phi_{\tau,\rho_E}(\rho))\le g(\tau N +(1-\tau)N_E),
\end{equation}
where $g(x):=(1+x)\log_2(1+x)-x\log_2 x$ and $N_E:=\text{Tr}\left(a^{\dagger}a\rho_E \right)$ is the mean photon number of the environment, which can be expressed as $({\text{Tr} \gamma_E \over 2}-1)/2$ for centered Gaussian states having the covariance matrix $\gamma_E$. In order to obtain a bound on the second term, we need to use \textsc{cQEPI}~\cite{cQEPI} expressed as
\begin{equation}
S(\rho_{X_1}\boxplus_\tau\rho_{X_2}|Z_1 Z_2)\ge\tau S(\rho_{X_1}|Z_1)+(1-\tau)S(\rho_{X_2}|Z_2),
\end{equation}
for all product states $ \rho_{X_1 Z_1} \otimes \rho_{ X_2 Z_2}$, where the conditional entropy $S(\rho_{X}|Z):=S(\rho_{XZ})-S(\rho_Z)$. In our case, the environment and output of complementary channel are conditioned by the purifying system, and the input and environment state is a product state by the definition of the channel. Consequently,
\begin{align}\label{ver1}
S( \Phi^{\text{c} \otimes n}_{\tau,\rho_E}(\rho_n))-nS(\rho_R) &
=S( \Phi^{\text{wc} \otimes n}_{\tau,\rho_E}(\rho_n)|R) \nonumber \\&
\ge (1-\tau) S(\rho_n)+\tau S(\rho^{\otimes n}_E|R) \nonumber \\ 
&=(1-\tau) S(\rho_n)-n\tau S(\rho_E) \nonumber \\
&\ge -n\tau S(\rho_E),
\end{align}
where the first inequality follows from the \textsc{cQEPI}, the second equality comes from independent and identically distributed (i.i.d.) assumption for environmental noise $\rho_E$ and $S(\rho_{ER})=0$, and the last inequality is obtained from the non-negativity of the entropy. Finally, we get the inequality as $S( \Phi^{\text{c} \otimes n}_{\tau,\rho_E}(\rho_n))\ge n(1-\tau)S(\rho_E)$. Note that if the environment is a Gaussian state, then $S(\rho_E)=g(N_\text{th})$, where $N_\text{th}$ is the mean thermal photon number of the environment, i.e., $\sum_i (\nu_i-1)/2$ for the symplectic eigenvalues $\nu_i$ of a given covariance matrix. For general cases, $N_\text{th} \equiv g^{-1}(S(\rho_E))$. Then Eq.~(\ref{Qcap}) becomes
\begin{align}
{\cal Q}(\Phi_{\tau,\rho_E},N)& \le g(\tau N +(1-\tau)N_E)-(1-\tau) S(\rho_E) \nonumber\\
&\equiv {\cal Q}_{U_1}.
\end{align}

Instead, we can consider the stronger \textsc{cQEPI} than the linear version, which is the exponential form given in Ref.~\cite{cQEPI2},
\begin{equation}
e^{S(\rho_{X_1}\boxplus_\tau\rho_{X_2}|Z_1Z_2)/n}\ge\tau e^{S(\rho_{X_1}|Z_1)/n}+(1-\tau)e^{S(\rho_{X_2}|Z_2)/n}.
\end{equation}
Then Eq.~(\ref{ver1}) can be modified as 
\begin{align}\label{ver2}
{1 \over n}S(\Phi^{\text{wc} \otimes n}_{\tau,\rho_E}(\rho_n)|R)&\ge \log\left( (1-\tau)e^{S(\rho_n)}+\tau e^{S(\rho^{\otimes n}_G|R)/n}\right) \nonumber \\
& \ge \log \left(1-\tau+\tau e^{-S(\rho_E)}\right).
\end{align}
Consequently, we have another upper bound such as ${\cal Q}(\Phi_{\tau,\rho_E},N) \le {\cal Q}_{U_2}$ where ${\cal Q}_{U_2} \equiv g(\tau N +(1-\tau)N_E)-\log \left( (1-\tau)+\tau e^{-S(\rho_E)} \right)-S(\rho_E)$.

For the amplifiers, the linear and exponential forms of \textsc{cQEPI} are also given in Ref.~\cite{cQEPI2} as follows.
\begin{align}
S(\rho_{X_1}&\boxplus_\kappa\rho_{X_2}|Z_1 Z_2)\ge {\kappa \over 2\kappa-1} S(\rho_{X_1}|Z_1) \nonumber
\\&+{\kappa-1 \over 2\kappa-1}S(\rho_{X_2}|Z_2)+\log{(2\kappa-1)},\;\;{\textnormal{and}}
\end{align}
\begin{equation}
e^{S(\rho_{X_1}\boxplus_\kappa\rho_{X_2}|Z_1Z_2)/n}\ge\kappa e^{S(\rho_{X_1}|Z_1)/n}+(\kappa-1)e^{S(\rho_{X_2}|Z_2)/n}, \label{amver2}
\end{equation}
where $\boxplus_\kappa$ is the two-mode squeezing operation, which corresponds to the amplifying parameter $\kappa \in [1,\infty]$. Then using the similar argument for the case of attenuator, we can obtain an upper bound for quantum capacity of the general amplifier channel ${\cal Q}(\Phi_{\kappa,\rho_E},N)$ from the linear \textsc{cQEPI} such as
\begin{align}
{\cal Q}(\Phi_{\kappa,\rho_E},N) &\le g(\kappa N+(\kappa-1)(N_E+1))
\nonumber \\
&-{\kappa-1 \over 2\kappa-1}S(\rho_E)-\log(2\kappa-1) \nonumber\\
&\equiv {\cal Q}_{U_1}^a.
 \end{align}
Similarly, we can also get ${\cal Q}_{U_2}^a$, which follows from Eq.~(\ref{amver2}),
\begin{align}
{\cal Q}_{U_2}^a &\equiv g(\kappa N+(\kappa-1)(N_E+1))
\nonumber \\&-\log \left( \kappa-1+\kappa e^{-S(\rho_E)} \right)-S(\rho_E).
\end{align}
It is worth mentioning that the upper bounds increase as the environment energy (average photon number $N_E$) increases. However, it doesn't mean actual quantum capacity always depends on the environment energy, e.g., coherent state environment.

\section{Lower bounds on the quantum capacity}\label{lower bounds}
Now, we need to consider proper lower bounds on the quantum capacity for our general attenuators and amplifiers in order to compare with the upper bounds. We can obtain lower bounds on those channels by means of Gaussian optimizer with fixed input entropy~\cite{palma17}, in which the thermal state reaches the minimum output entropy of the given channel. We can express a lower bound of the quantum capacity for the general attenuator as
\begin{align}\label{lower}
{\cal Q}(\Phi_{\tau,\rho_E},N) &\ge  \max_{\bar{E}(\rho) \le N} I_c\left( \Phi_{\tau,\rho_E},\rho \right)
\nonumber \\ &\ge S(\Phi_{\tau,\rho_E}(\rho_{\text{th},N}))-S(\Phi^\text{c}_{\tau,\rho_E}(\rho_{\text{th},N})),
\end{align}
where the second inequality from using a specific thermal state as an input state, instead of optimizing over all possible states. In order to obtain a bound on the first term, we recall $\Phi_{\tau,\rho_E}(\rho_{\text{th},N})=\Phi_{1-\tau,\rho_{\text{th},N}}(\rho_E)$ by considering the corresponding characteristic functions~\cite{KSIEEE}. Then, by the fact that the output entropy of the single mode phase-insensitive Gaussian channel for a fixed input entropy is minimized by the thermal state having the same entropy, we can get the inequality~\cite{palma17}, 
\begin{align}
S(\Phi_{\tau,\rho_E}(\rho_{\text{th},N}))&=S(\Phi_{1-\tau,\rho_{\text{th},N}}(\rho_E)) \nonumber\\
&\ge g((1-\tau)N_\text{th}+\tau N).
\end{align}
For a bound on the second term of Eq.~(\ref{lower}), we use the maximality of Gaussian state again as in Ref.~\cite{Wolf}, then
\begin{align}
S(\Phi^\text{c}_{\tau,\rho_E}(\rho_{\text{th},N}))& \le S(\Phi^\text{wc}_{\tau,\rho_E}(\rho_{\text{th},N}))+S(\rho_R)
\nonumber \\ &\le g((1-\tau)N+\tau N_E))+S(\rho_E),
\end{align} 
where $R$ is the reference system for purifying environment and using the fact that $\Phi^\text{wc}_{\tau,\rho_E}=\Phi_{1-\tau,\rho_E}$. Finally we get the lower bound on the quantum capacity for the general attenuator as
\begin{align}
{\cal Q}(\Phi_{\tau,\rho_E},N) &\ge g((1-\tau)N_\text{th}+\tau N) \nonumber\\
&-g((1-\tau)N+\tau N_E)-S(\rho_E) \nonumber\\
&\equiv {\cal Q}_{L}.
\end{align}

Similarly, a lower bound on the quantum capacity of the amplifier can be written as
\begin{align}\label{alower}
{\cal Q}(\Phi_{\kappa,\rho_E},N) &\ge  \max_{\bar{E}(\rho) \le N} I_c\left( \Phi_{\kappa,\rho_E},\rho \right)
\nonumber \\ &\ge S(\Phi_{\kappa,\rho_E}(\rho_{\text{th},N}))-S(\Phi^\text{c}_{\kappa,\rho_E}(\rho_{\text{th},N})).
\end{align}
The first term is bounded from below as in Ref.~\cite{palma17}, and so we have 
\begin{align}
 S(\Phi_{\kappa,\rho_E}(\rho_{\text{th},N}))&=S(\Phi_{\kappa-1,\rho_{\text{th},N}}(\rho_E)) 
 \nonumber \\ &\ge g((\kappa-1)N_\text{th}+\kappa (N+1)),
\end{align}
and the second term is bounded from above using maximality of Gaussian state as in Ref.~\cite{Wolf}, and so we obtain
\begin{align}
S(\Phi^\text{c}_{\kappa,\rho_E}(\rho_{\text{th},N}))&\le S(\Phi^\text{wc}_{\kappa,\rho_E}(\rho_{\text{th},N}))+S(\rho_R) 
\nonumber \\ &\le g((\kappa-1)N+\kappa (N_E+1))+S(\rho_E).
\end{align}
Consequently, we get the lower bound on the quantum capacity of amplifier as follows:
\begin{align}
{\cal Q}(\Phi_{\kappa,\rho_E},N) &\ge g((\kappa-1)N_\text{th}+\kappa (N+1))
\nonumber \\ &-g((\kappa-1)N+\kappa (N_E+1))-S(\rho_E)
\nonumber \\ &\equiv {\cal Q}_{L}^a.
\end{align}

\begin{figure}[!t]
\centering
\includegraphics[width=8.6cm]{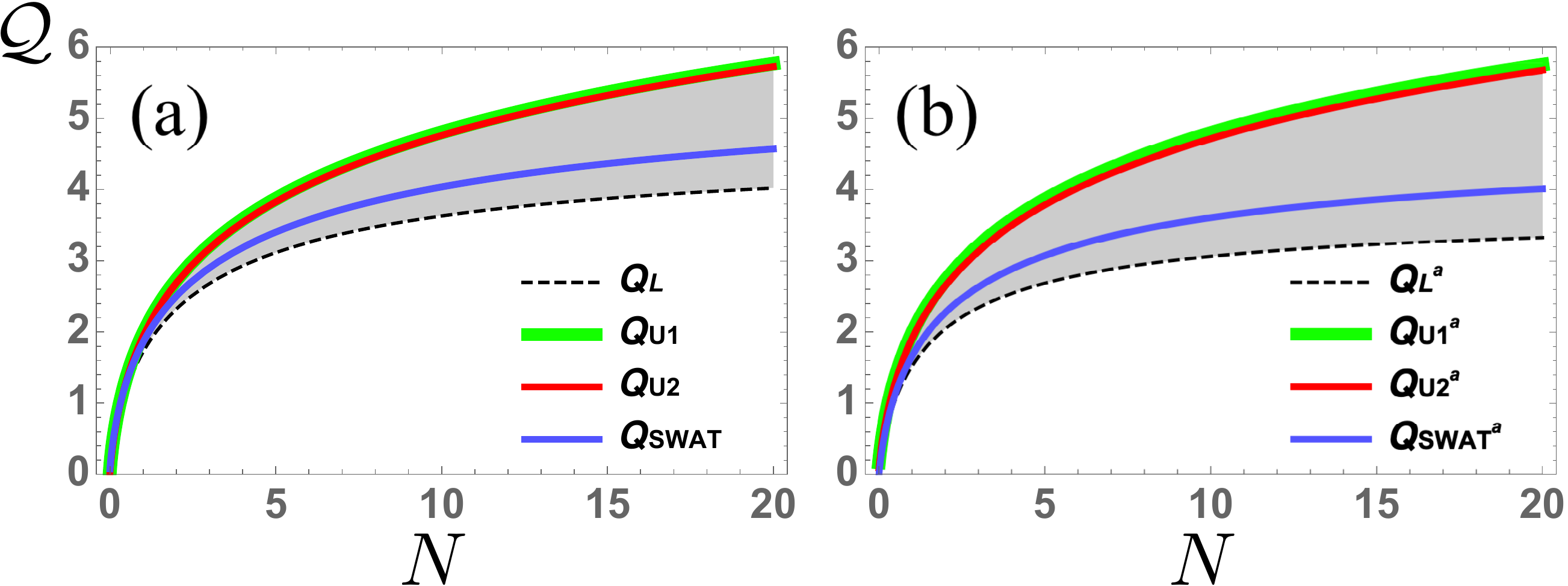}
\caption{Comparison between our two upper bounds with known upper bound $Q_\text{SWAT}$~\cite{constrained} for (a) the thermal attenuator with $\tau=0.99$ and (b) the amplifier with $\kappa=1.02$. Average photon number of the thermal environment is $N_E=N_\text{th}=1$.}
\label{fig2}
\end{figure}

\section{examples}\label{examples}
In the previous sections, we have investigated our upper and lower bounds on the quantum capacity for the case in which environment can be any state in general. Here we give specific examples in order to consider the physical meanings of our results. The first nontrivial example whose quantum capacity is unknown is the thermal attenuator, in which the environment is the thermal state. Unfortunately, our upper bounds cannot improve them (See Fig.~\ref{fig2}). Therefore we investigate more general Gaussian environment, i.e., squeezed thermal state, to the non-Gaussian environment. 
\begin{figure}[!t]
\centering
\includegraphics[width=8.6cm]{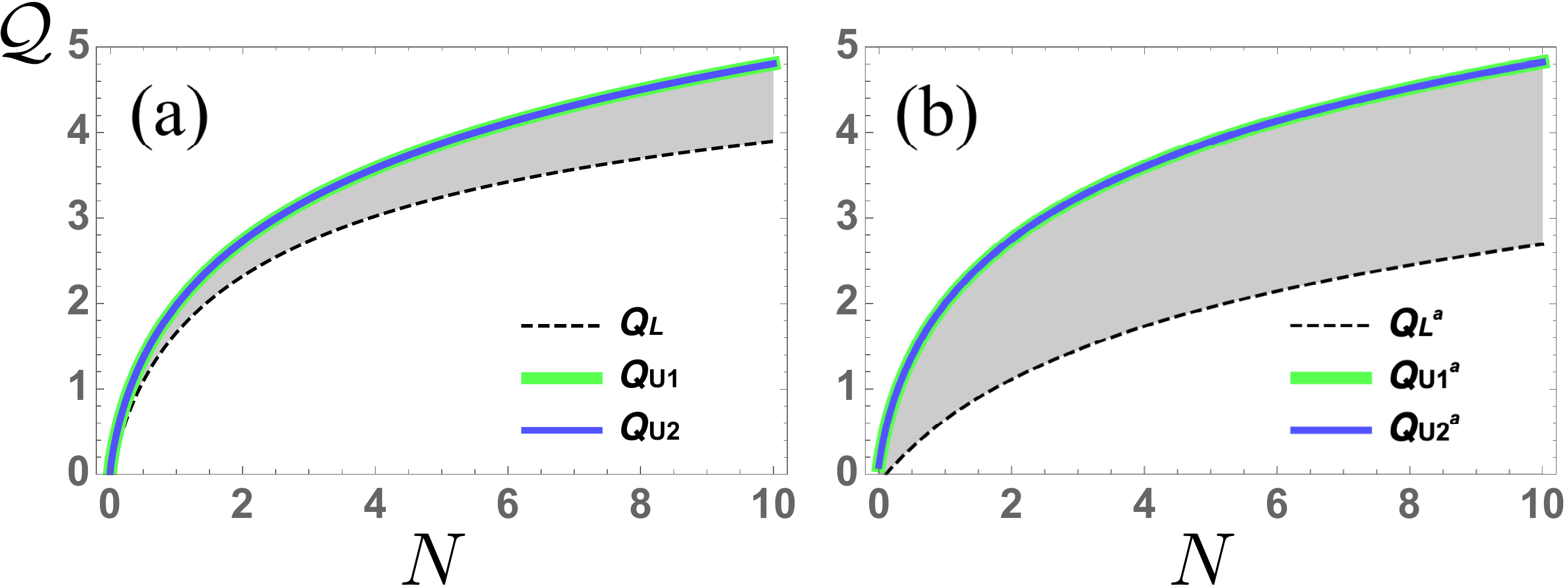}
\caption{Two upper bounds and a lower bound on the quantum capacity of Gaussian (a) attenuator with $\tau=0.98$ and (b) amplifier with $\kappa=1.02$ on the squeezed thermal environment, when $N_\text{th}=0.01$ and squeezing parameter $r=0.1$, thus $N_E \sim0.02$. $N$ is the mean photon number of input state, ${\cal Q}$ is the quantum capacity (bits). Note that the two upper bounds are very close; thus they are overlapped in both cases.}
\label{fig3}
\end{figure}

\subsection{Squeezed thermal environment}
We can express the covariance matrix of a centered squeezed thermal state as
\begin{equation}
\gamma_\text{sth}=(2N_\text{th}+1)\begin{pmatrix} e^{2r} & 0 \\ 0 & e^{-2r} \end{pmatrix},
\end{equation}
where $N_\text{th}$ is the mean photon number from the thermal noise, and $r \in [0,\infty)$ is the squeezing parameter. Then the mean photon number $N_E$ of this state can be written as
\begin{equation}
N_E={1 \over 2}\left({\text{Tr} \gamma_\text{sth} \over 2}-1\right)={1 \over 2}\big((2N_\text{th}+1)\cosh{2r}-1\big).
\end{equation}
Therefore, we can easily obtain the value of entropy $S(\rho_E)=g(N_\text{th})$, and $N_E$ for given mean thermal photon $N_\text{th}$ and squeezing parameter $r$. The squeezed thermal state is the most general single-mode Gaussian state when its mean is placed at origin, which can be always removed by the local symplectic unitary transformation. Consequently, what we are considering here are general Gaussian attenuator and amplifier. We plot the upper and lower bounds of the quantum capacity with respect to input state energy in Fig.~\ref{fig3}. Our results give narrow bounds near the region of $\tau,\kappa \sim 1$, when the input energies are low.

\begin{figure}[!t]
\centering
\includegraphics[width=8.6cm]{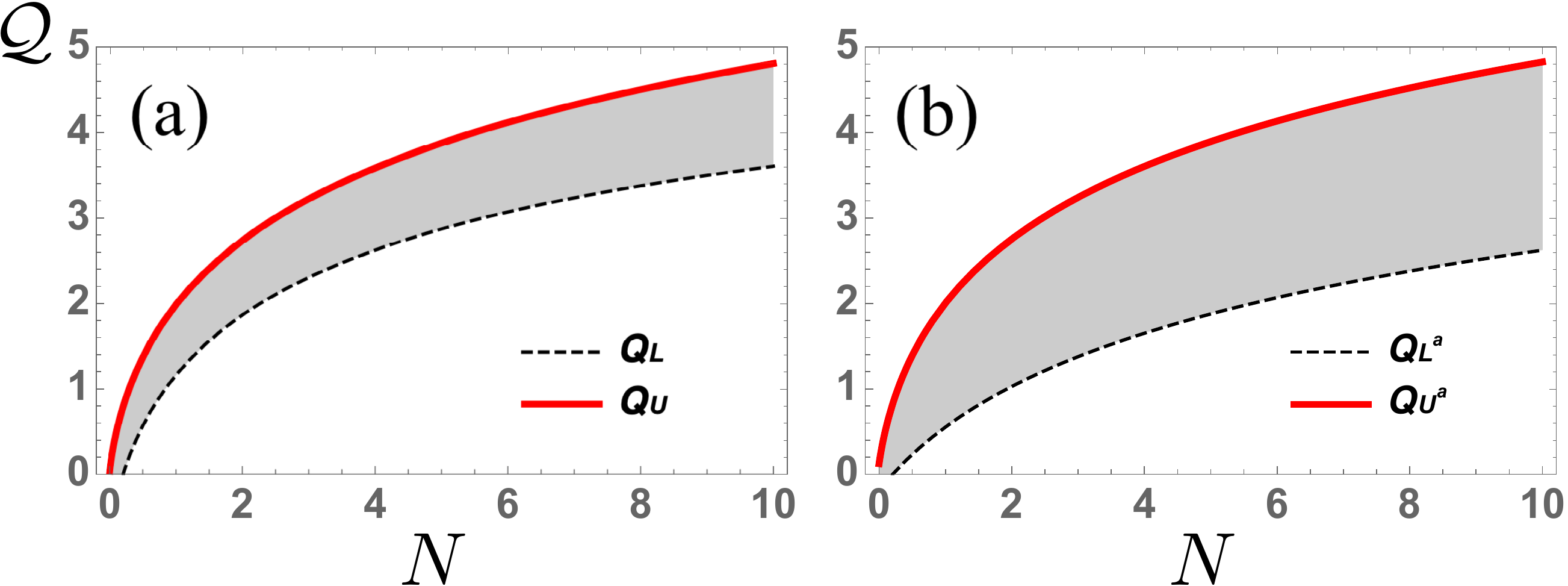}
\caption{Upper and lower bounds on the quantum capacity of general (a) attenuator with $\tau=0.98$ and (b) amplifier with $\kappa=1.02$. The mean photon number of the environment is $N_E=0.2$.}
\label{fig4}
\end{figure}

\subsection{Non-Gaussian environment}
As next examples, we investigate a pure non-Gaussian environment (e.g., Fock state) and a general mixed state. In the case of pure environment, $S(\rho_E)=N_\text{th}=0$ by definition. Therefore the upper and lower bounds have very simple forms such as
\begin{align}
&{\cal Q}_{U_1}={\cal Q}_{U_2}=g(\tau N +(1-\tau)N_E), \nonumber \\
&{\cal Q}_{L}=g(\tau N)-g((1-\tau)N+\tau N_E), \nonumber \\
&{\cal Q}_{U_1}^a={\cal Q}_{U_2}^a=g(\kappa N+(\kappa-1)(N_E+1))-\log(2\kappa-1),  \nonumber\\
&{\cal Q}_{L}^a=g(\kappa (N+1))-g((\kappa-1)N+\kappa (N_E+1)).
\end{align}
In Fig.~\ref{fig4}, we plot the upper and lower bounds for these channels. 

\begin{figure}[!t]
\centering
\includegraphics[width=8.6cm]{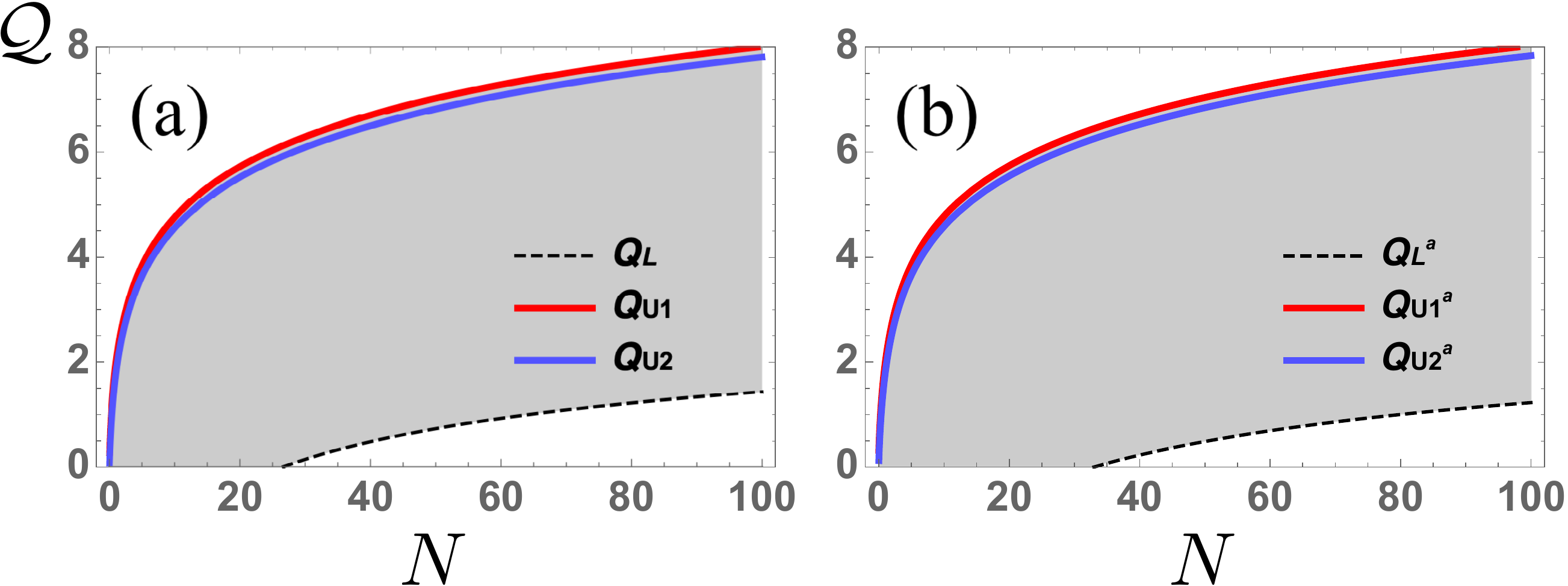}
\caption{Upper and lower bounds on the quantum capacity of general (a) attenuator with $\tau=0.98$ and (b) amplifier with $\kappa=1.02$. Entropy of the environment is $S(\rho_E)\sim 2.75$, thus $N_\text{th}=2$ and set $N_E=3$.}
\label{fig5}
\end{figure}

For the last example, we consider more noisy non-Gaussian environment, in a sense that the mean photon number and entropy are relatively high. We can figure out that our bounds are not so tight in this case and the two upper bounds split (Fig.~\ref{fig5}).

\section{discussions}\label{discussion}
We have investigated upper and lower bounds on the energy-constrained quantum capacity for general attenuator and amplifier. Our primary method is \textsc{cQEPI}, which can be used for obtaining bounds on the output entropy of the complementary channel. Although our results do not give tighter bounds over known results for thermal attenuator and amplifier, it is applicable to more general environment, no matter whether it is Gaussian or not. Moreover, we have shown that our bounds become tight ones when the channel transmissivity is near unity and the input energy is low.

Since the general attenuator and amplifier cannot cover all single-mode Gaussian channels, one of the most important works is finding an equivalent class of all single-mode Gaussian channels having the same quantum capacity. Furthermore, there is still a possibility for finding a tighter bound on the quantum capacity of the amplifier, as can be seen from our results (Fig.~\ref{fig3} (b) and Fig.~\ref{fig4} (b)). Even for the thermal amplifier, the known upper bound is not that tight~\cite{plob} compared with the thermal attenuator, so we have not observed any activation of the quantum capacity, which was investigated in Ref.~\cite{Lim19}. With these considerations, we expect that our work could extend the knowledge of the quantum capacity, which is still far from being fully understood.

\section*{ACKNOWLEDGMENTS}
This work was supported by Basic Science Research Program through the National Research Foundation of Korea (NRF) funded by the Ministry of Science and ICT (NRF-2016R1A2B4014928 \& NRF-2017R1E1A1A03070510). S. L. acknowledges the Ministry of Science and ICT, Korea, under an ITRC Program, IITP-2019-2018-0-01402, and K. J. acknowledges the Ministry of Education, Korea (NRF-2018R1D1A1B07047512).

\end{document}